\documentclass[twocolumn,showpacs,preprintnumbers,prl,fleqn,floatfix]{revtex4}

\usepackage{graphicx}
\usepackage{dcolumn}
\usepackage{bm}
\usepackage{amsmath, amssymb}

\begin{document}

\title{Fractional quantum Hall effect in CdTe}

\author{B. A. Piot$^{1}$, J. Kunc$^{1,2}$, M. Potemski$^{1}$, D. K. Maude$^{1}$, C. Betthausen$^{3}$, A. Vogl$^{3}$, D. Weiss$^{3}$, G. Karczewski$^{4}$, T. Wojtowicz$^{4}$}

\affiliation{$^{1}$ Laboratoire National des Champs Magn\'etiques
Intenses, CNRS-UJF-UPS-INSA, F-38042 Grenoble, France}

\affiliation{$^{2}$ Institute of Physics, Faculty of Mathematics
and Physics, 12116 Prague, Czech Republic}

\affiliation{$^{3}$ Departement of Physics, Regensburg University,
D-93053 Regensburg, Germany}

\affiliation{$^{4}$ Institute of Physics, Polish Academy of
Sciences, PL-02668 Warsaw, Poland}

\date{\today }

\begin{abstract}

The fractional quantum Hall (FQH) effect is reported in a high
mobility CdTe quantum well at mK temperatures. Fully-developed FQH
states are observed at filling factor $4/3$ and $5/3$ and are
found to be both spin-polarized ground state for which the lowest
energy excitation is not a spin-flip. This can be accounted for by
the relatively high intrinsic Zeeman energy in this single valley
2D electron gas. FQH minima are also observed in the first excited
(N=1) Landau level at filling factor $7/3$ and $8/3$ for
intermediate temperatures.

\end{abstract}
 \maketitle

Interacting carriers in certain FQH ground states can have
reversed spins provided the Zeeman energy is sufficiently small.
This is typically observed in GaAs-based 2D electron gases
(2DEGs), where an increase in the Zeeman energy induces a change
in the spin polarization of the ground state from unpolarized to
fully spin-polarized. This transition has been reported for the
FQH states at filling factor $\nu$=4/3, $\nu$=8/5, $\nu$=2/3, or
$\nu$=2/5 \cite{Clark89,Eisentein89,Engel92,Kang97}, as well as in
a GaAs 2D hole gas \cite{Davies91}. Subsequently, this behavior
was elegantly interpreted within the composite fermions (CF) model
\cite{Jain89} for the FQH effect by invoking Zeeman energy-induced
crossings between spin-split composite fermion Landau levels,
leading to possible changes of the spin configuration of the
ground state \cite{Du95}. More recently, the $\nu$=4/3 FQH state
was investigated in a strained Si quantum well \cite{Lai04}, where
the associated resistance minimum was found to maintain its
strength with increasing Zeeman energy, which was interpreted as
the consequence of a spin-polarized ground state. The latter work
addresses the interesting question of how the FQH effect manifests
itself in a 2D system with an intrinsically larger Zeeman energy
than in GaAs. However, the influence of the valley degeneracy
inherent in Si is another degree of freedom that may also
interfere with the FQH physics.

In the present work, we study the evolution of FQH states under
relatively high intrinsic Zeeman energy in a \textit{single
valley} electron system. This is made possible by investigating
the FQH effect in a high quality 2D electron gas in CdTe, a single
valley, direct gap, semiconductor in which the bare electronic
g-factor is about four times larger than in GaAs. A fundamental
asset of this system is the possibility to incorporate magnetic
ions to form a so-called diluted magnetic semiconductor, which
offers possible applications in the fields of spintronics and
quantum computing. The transport measurements performed at mK
temperature reveal \emph{fully-developed} FQH states (i.e. zero
longitudinal resistance and exact quantization of the Hall
resistance) in the upper spin branch of the lowest (N=0) Landau
level (LL), which constitutes to our knowledge the first
observation of the FQH effect in a II-VI semiconductor. Tilted
magnetic fields experiments up to 28 Tesla show no significant
changes of the FQH gap both at filling factor $4/3$ and $5/3$, a
behavior typical of spin-polarized ground states for which the
lowest energy excitation is not a spin-flip. This can be accounted
for by the relatively high intrinsic Zeeman energy which wins over
the Coulomb energy to force the spins to align with the magnetic
field. This can also be seen as the consequence of energy level
crossings in the composite fermion approach for the FQH effect.
Significantly, emerging FQH minima at filling factor $7/3$ and
$8/3$ are also observed at intermediate temperatures in the first
excited (N=1) LL, demonstrating the high quality of the 2DEG that
it is now possible to achieve in this material. This leaves open a
possible future observation of the $\nu$=5/2 FQH state in the
presence of a relatively high intrinsic Zeeman energy.

The sample studied here is a 20 nm-wide CdTe quantum well,
modulation doped on one side with iodine, and embedded between
Cd$_{1-x}$Mg$_{x}$Te barriers ($x\simeq$0.26). It was cooled down
in a $^{3}He/^{4}He$ dilution fridge to mK temperature in a number
of different ways: under continuous illumination with a green
laser or a green light emitting diode (LED), under continuous
illumination with a yellow LED, and, in the darkness. These types
of cooldowns will be referred to as cooldown A, B and C
respectively. The resulting electron density for cooldown A, B,
and C, are $n_{s}=4.50, 4.53$ and $3.80\times 10^{11}cm^{-2}$
respectively, and the electron mobility at T$\sim600$mK for
cooldown A is around $\mu=260 000 cm^{2}/V s$. Transport
measurements were performed with a standard low frequency lock-in
technique for temperatures between 40 mK and 1.4K under magnetic
fields up to 28 T.

\begin{figure}[tbp]
\includegraphics[width=1\linewidth,angle=0,clip]{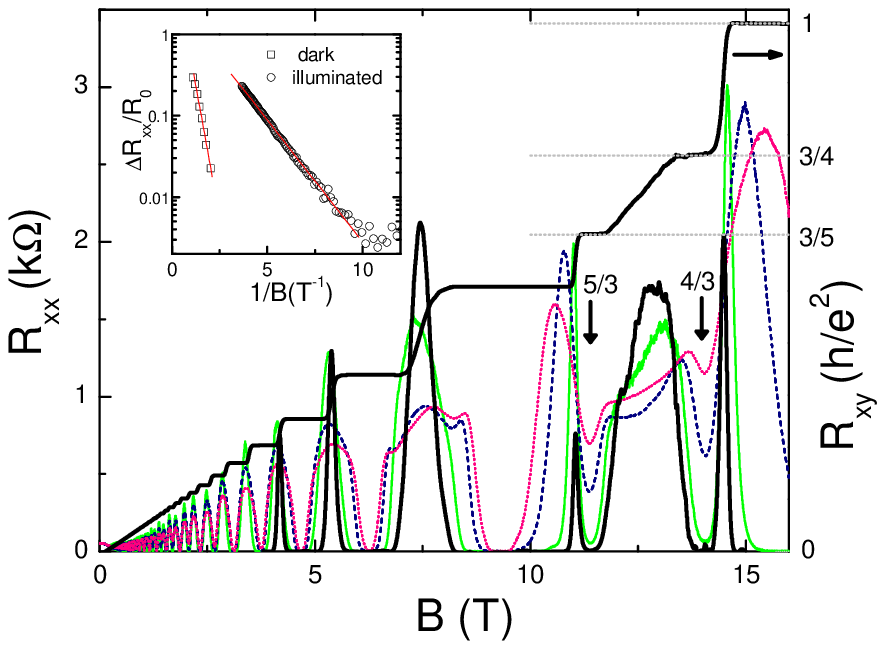}
\caption{(Color online) Hall resistance $R_{xy}$ and longitudinal
resistance $R_{xx}$ versus perpendicular magnetic field for
different temperatures, $T=40$mK (thick solid line), $T=90$mK
(thin solid line), $T=534$mK (dashed line), and $T=1.03$K (dotted
line). Cooldown A. Inset: Dingle plots. Semi-logarithmic plot of
$\Delta R_{xx}/R_{0}=(R_{xx}-R_{0})/R_{0}$ where $R_{0}$ is the
zero-field resistance versus $1/B$ for cooldown A (open circles)
and for a cooldown C (open squares).}\label{fig1}
\end{figure}

In Fig.\ref{fig1}, we plot the longitudinal resistance $R_{xx}$
for cooldown A as a function of the perpendicular magnetic field
for different temperatures.
Pronounced FQH states are observed at low temperature at filling
factor $\nu$=4/3, and $\nu$=5/3, with the resistance falling to
zero, together with well defined quantized Hall resistance. The
role of illumination in improving the sample quality is critical,
as can be seen in the inset of Fig.\ref{fig1} where we plot the
amplitude of the low field Shubnikov de Haas (SdH) oscillations as
a funtion of the inverse magnetic field on a logarithmic scale at
T$\sim90$mK, before and after illumination. The so-called Dingle
plot exhibits a linear decay whose slope $\pi m^{*}/e\tau_{q}$,
where $m^{*}$ is the electron effective mass, gives an estimation
of the electron quantum lifetime $\tau_{q}$
\cite{Coleridge96,PiotPRBsto}. While the transport lifetime
$\tau_{tr}=\mu m^{*}/e$ increases by a factor of 2 after
illumination, the quantum lifetime is found to increase more than
five times, jumping from $0.6$ ps in the dark to $3\pm0.3$ ps
after illumination, confirming the considerable improvement of the
sample quality. This value of $\tau_{q}$ is comparable to the one
which can be observed in GaAs samples with several million
mobility, despite our moderate measured mobility of $260 000
cm^{2}/V s$. This apparent contradiction is basically due to the
fact that in such high mobility GaAs samples, the long-range
scattering by remote donors is even more predominant and leads to
a longer transport time $\tau_{tr}$, for a comparable $\tau_{q}$.

Nevertheless, at low temperature, an important number of
electronic states are localized, leading to wide zero resistance
states in the integer quantum Hall effect which prevent the
observation of any signs of the FQH effect in the first excited
(N=1) LL. As the temperature is increased, the fraction of
localized states is reduced and weak FQH minima become visible in
the N=1 LL. These features persist up to relatively high
temperature, demonstrating again the quality of the sample.

\begin{figure}[tbp]
\includegraphics[width=1\linewidth,angle=0,clip]{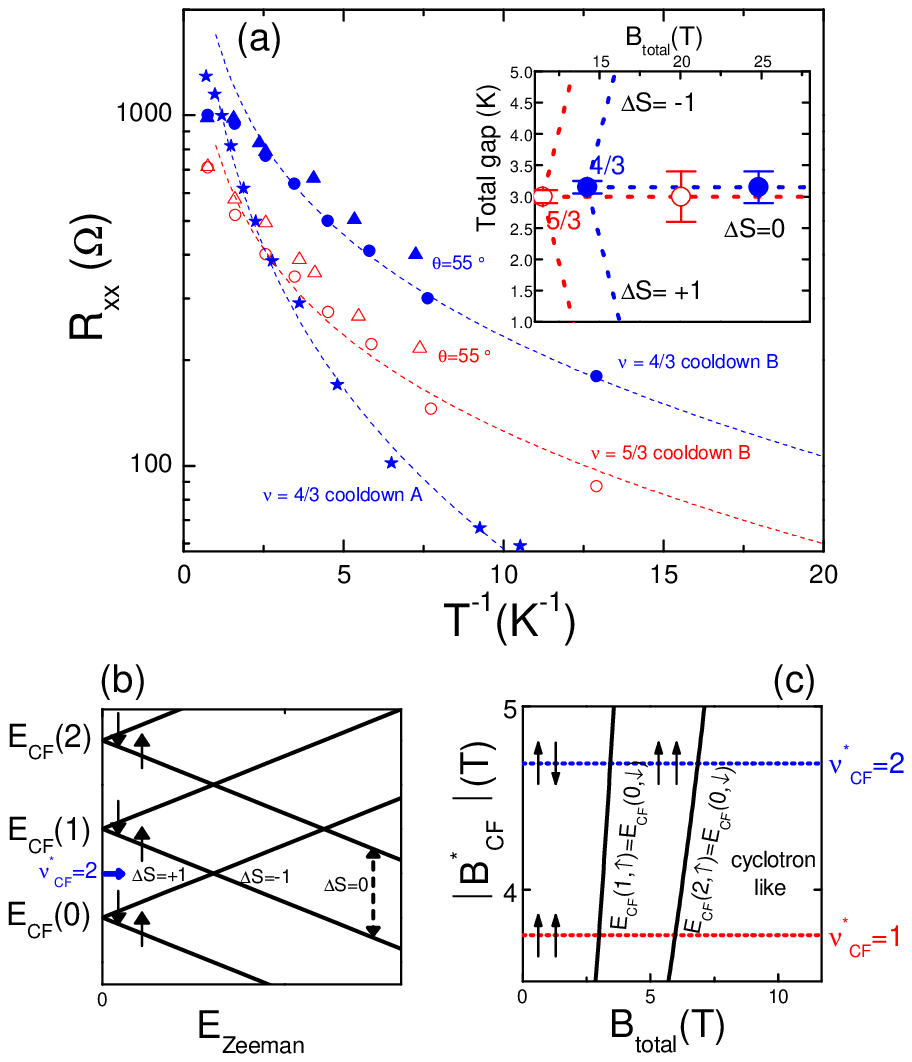}
\caption{(Color online) (a) Longitudinal resistance $R_{xx}$ at
$\nu$=4/3 as a function of inverse temperature for cooldown A
(stars) and for cooldown B at $\theta=0^{\circ}$ (circles) and
$\theta=55.6^{\circ}$ (triangles). Same data for cooldown B at
$\nu$=5/3 (open symbols). Simulations of the thermally activated
resistance (dashed lines) (see text). Inset: Corresponding total
FQH gaps at $\nu$=4/3 and $\nu$=5/3 as a function of the total
field $B_{total}$. Expected evolution of the gaps for different
ground states (dotted lines) (see text). (b) Schematic
representation of the CF fan diagram at fixed CF cyclotron energy,
as a function of the Zeeman energy (see text). $E_{CF}(N)$ is the
energy of the N$^{th}$ CF level. The arrows depict the spin
orientation of each sub-band. (c) Position of the CF level
crossings in the ($B_{CF}^{*}$,$B_{total}$) plane (see text). The
arrows depict the spin polarization of the ground state in
different region.}\label{fig2}
\end{figure}

In Fig.\ref{fig2} we focus on the FQH effect in the N=0 Landau
level. Fig.\ref{fig2}.a. shows the temperature dependance of the
longitudinal resistance at $\nu$=5/3 and 4/3, for cooldown A and B
as a function of the inverse temperature. The difference in sample
quality between cooldowns appears clearly when comparing the low
temperature behavior of the initially similar resistance at
filling factor $4/3$. The so-called ``activation plots'' or
Arrhenius plots are generally used to extract an activation gap or
mobility gap, corresponding to the energy difference between the
edge of the delocalized states of the ground and excited state. In
the FQH regime, this activation gap has been shown to be in
agreement with the calculated FQH mobility gap once disorder (and
other corrections) are taken into account (see e.g.
Refs.~[\onlinecite{ZhangDasSarma86,Wan05}]). However, a simple
extraction of this activation gap $\Delta$ requires the
observation of an expanded linear region (typically at least one
order of magnitude) where $R_{xx} \sim
e^{\frac{-\Delta}{2k_{B}T}}$, whereas such a region is rather
absent in our data. This none thermally-activated behavior is
actually expected when an accurate level shape is taken into
account (i.e. Gaussian or Lorentzian broadening), for which one
expects the linear behavior in an activation plot to deviate at
low temperature in the presence of the broadening which reduces
the mobility gap. This effect becomes important when the particles
level broadening is non-negligible compared to the total
(spectral) gap. To analyze our data, we therefore use the model
proposed in Ref.~[\onlinecite{Usher91}] which includes a
disorder-induced Gaussian broadening to calculate the temperature
dependence of the resistance. The results of these simulations
(which details will be given elsewhere) are plotted as
dotted-lines in Fig.\ref{fig2}.a and show a very good agreement
with the experimental behavior.

Activation data was also collected when tilting the 2DEG plane in
the total magnetic field with an in-situ rotation stage at an
angle of $\theta=55^{\circ}$. This data, also plotted in
Fig.\ref{fig2}.a, is very similar to the $\theta=0^{\circ}$
behavior for $\nu$=4/3 and $\nu$=5/3. The small difference can be
well reproduced for both fractions either by introducing a small
increase ($\sim 10 \%$) in the level width, while the total gap
remains constant, or by using a constant level width and a
slightly reduced gap ($\sim 10 \%$ also). The total gap extracted
from our analysis at $\theta=0^{\circ}$ and $\theta=55^{\circ}$
are plotted in the inset of Fig.\ref{fig2}.a. as a function of the
total field at fixed perpendicular field (filling factor), the
vertical error bar representing the possible gap decrease at
$\theta=55^{\circ}$. We also plot here the expected evolution of
the total gap as a function of total magnetic field (Zeeman
energy) in three different configurations: a spin-polarized ground
state with single particle spin-reversed excitation ($\Delta
S=-1$, where $\Delta S$ is the net spin change of the excitation),
a spin-polarized ground state with no spin reversed excitations
($\Delta S=0$), and a spin-unpolarized ground state ($\Delta
S=+1$). The bare g-factor $g^{*}=-1.6$ is taken from Raman
scattering measurements performed on the same sample.

The fact that the $\nu$=5/3 gap remains nearly constant at
$\theta=55^{\circ}$ suggests, as observed in GaAs, a
spin-polarized ground state with a lowest energy excitation which
is not a spin-flip, since no increase is observed despite of a
significant variation (nearly a factor of 2) of the Zeeman energy.
If the $\nu$=4/3 state was to be unpolarised, one would expect a
sharp decrease of the gap as well as its disappearance, here
around $B_{total}=$16T , before reentrance at higher fields due to
a change in the ground state polarization. This transition has
been observed in GaAs 2DEG at low electron density
\cite{Clark89,Du95}, and also for higher densities close to the
one of our CdTe sample. In
Refs.~[\onlinecite{Clark89,Clarksurf88}], the $\nu$=4/3 FQH gap
for sample G71 with initial electron density $\sim2.7\times
10^{11}cm^{-2}$ decreases as the density (total field) is
increased and is close to vanishing for magnetic fields of about
12 T. Our observation of a quasi-unchanged gap at
$\theta=55^{\circ}$ shows the $\nu$=4/3 FQH state is
spin-polarized in CdTe. The fact that this gap is \textit{not
increasing} further suggests that the lowest energy excitations in
this state do not involve spin reversal.

The qualitative behavior of the gap at different tilt angles
between $\theta=0^{\circ}$ and $\theta=55^{\circ}$ can be inferred
from a detailed angular dependence of $R_{xx}$ measured for a
fixed intermediate temperature of $T\sim 390$mK. At this
temperature the gap variation can efficiently be probed as
observed when comparing the resistance values at $\nu$=4/3 and
$\nu$=5/3 for cooldown A and B (Fig.\ref{fig2}.a.). This angular
dependence (not shown) shows only a very weak variation of the
resistance at $\nu$=5/3 and $\nu$=4/3 over the entire $\theta$
range studied ($0<\theta<55^{\circ}$). This confirms that no
significant changes in the $\nu$=4/3 and $\nu$=5/3 FQH gaps are
observed upon tilting, as expected for a spin-polarized state with
no spin-reversed excitation.

This behavior can actually be understood more quantitatively using
the CF theory for FQH effect, where FQH for electron is mapped
onto the integer quantum Hall effect for composite fermions. In
the upper spin branch of the $N=0$ LL, around $\nu$=3/2, these CF
see an effective magnetic field
$B_{CF}^{*}=3(B_{\bot}-B_{\bot3/2})$, where $B_{\bot3/2}$ is the
magnetic field corresponding to $\nu$=3/2 \cite{Halperin92,Du95}.
In this case the $\nu$=4/3(5/3) FQH effect for electrons is the
$\nu^{*}_{CF}=2(1)$ integer quantum Hall effect for CF. The scale
of the CF cyclotron gap between two CF levels is then given by
$\hbar e B_{CF}^{*}/m_{CF}^{*}$, where $m_{CF}^{*}$ is the CF
effective mass. When the Zeeman energy is added to this simple
picture, which is schematically depicted in Fig.\ref{fig2}.b, the
lower spin branch of the N=1 CF level ($(1,\uparrow)$) may have a
lower energy than the upper spin branch of the N=0 CF level
($(0,\downarrow)$). In this situation the ground state at
$\nu^{*}_{CF}=2$, initially formed by $(0,\uparrow)$ and
$(0,\downarrow)$ CF levels for small Zeeman energies, is now
formed by the $(1,\uparrow)$ and $(0,\uparrow)$ CF levels and
therefore spin-polarized. This picture can be applied to our 2DEG
in CdTe, with a g-factor of $g^{*}=-1.6$ and the composite
fermions effective mass experimentally determined in
Ref.~[\onlinecite{Leadley94}] as a function of $B_{CF}^{*}$
($m_{CF}^{*}=0.51+0.074 B_{CF}^{*}$). In Fig.\ref{fig2}.c., we
plot in a ($B_{CF}^{*}$,$B_{total}$) plane the position of the
crossing points of the $(0,\downarrow)$ CF level with the
$(1,\uparrow)$ and $(2,\uparrow)$ levels. For
$\nu=4/3(\nu^{*}_{CF}=2)$, these crossings occur for
$B_{total}\sim3.4$T and $B_{total}\sim6.8$T respectively,
explaining why the $\nu$=4/3 FQH ground state is spin-polarized
with no spin-reversed excitations for the total magnetic field
range investigated ($14<B_{total}<25$T). The excitation gap in
this domain corresponds  to a CF cyclotron gap (referred to as
``cyclotron-like'' in Fig.\ref{fig2}.c.). The same conclusions are
drawn for the $\nu=5/3(\nu^{*}_{CF}=1)$ FQH state, provided
$B_{total}>3$T. We note that the CF cyclotron gap used in these
calculations is larger than the experimentally measured FQH gap
discussed above, meaning that the transition to ``cyclotron-like''
excitations should occur at even smaller magnetic field.

\begin{figure}[tbp]
\includegraphics[width=1\linewidth,angle=0,clip]{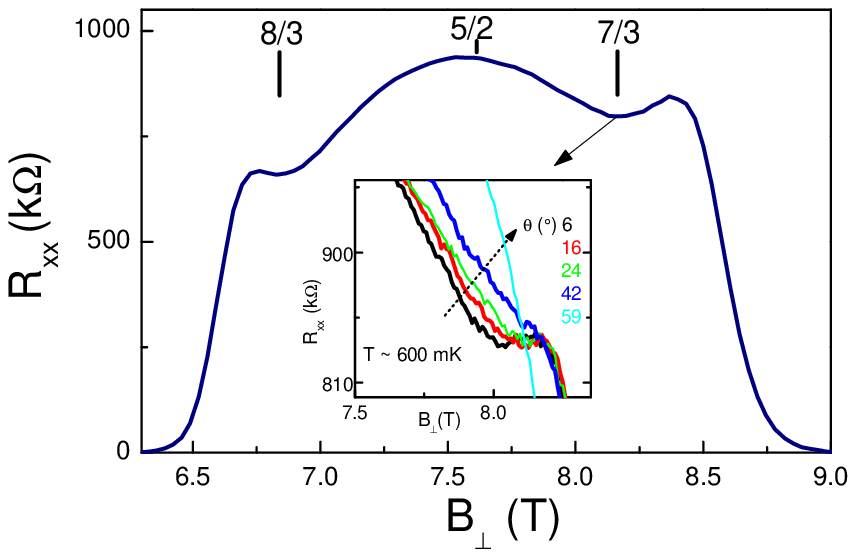}
\caption{(Color online) Magneto resistance $R_{xx}$ in the lower
spin branch of the N=1 LL at $T=534$mK. Corresponding filling
factors are indicated on top. Inset: angular dependance of the
$\nu$=7/3 minimum at $T=600$mK. Increasing tilting angle $\theta$
indicated by the arrow.}\label{fig3}
\end{figure}

Finally, we turn to the description of the emerging FQH effect in
the $N=$1 LL which can be observed in our sample at intermediate
temperatures. As can be seen in Fig.\ref{fig1}, weak minima are
emerging at filling factors $\nu$=7/3 and $\nu$=8/3 for
temperatures above 400-500 mK. At lower temperatures, the
increasing number of localized states leads to the FQH effect
being masked by the integer quantum Hall effect. The $T=534$ mK
perpendicular field data of Fig.\ref{fig1} are replotted for
clarity in Fig.\ref{fig3}. We note that no minimum is observed at
filling factor $\nu$=5/2, most likely due to insufficient sample
quality. The behavior of the $N=$1 FQH effect under tilted
magnetic field have recently been revisited
\cite{Lilly99,Pan99,Dean08} notably due to the extraordinary
interest in the even-denominator $\nu$=5/2 FQH state (for a
review, see Ref.~[\onlinecite{RevModPhys.80.1083}]). The
tilted-field behavior of the $\nu$=7/3 gap turns out to be
non-trivial, increasing with tilt in the low field limit
\cite{Dean08}, decreasing in higher density samples
\cite{Eisensteinsurfsci}, and eventually disappearing at high
angles where an anisotropic phase settles \cite{Lilly99,Pan99}.
The tilted-field behavior of the magnetoresistance in the N=1 LL
was examined at intermediate temperature in our CdTe 2DEG and is
presented in Fig.\ref{fig3}. In the inset, we focus on the
evolution of the local minimum at $\nu$=7/3 and $T=600$mK for
different tilt angles. We observe that this minimum maintains its
strength at low angles, before starting to weaken around
$\theta=24^{\circ}$ and finally disappearing for
$\theta>42^{\circ}$. The relative initial stability with respect
to tilt angle is similar to the one observed in the N=0 LL, and
suggest that, as for $\nu$=5/3 and $\nu$=4/3, the $\nu$=7/3 state
is already in a regime where the ground state is spin-polarized
with a lowest energy excitation which is not a spin flip. However,
the observation of a $\nu$=7/3 state at lower temperatures (not
possible because of localization) would be necessary to validate
this hypothesis. At higher angles however, the minimum clearly
disappears and the resistance at the broad maximum in $R_{xx}$
associated with the $N=1$ LL starts to increase. Depending on the
orientation between the parallel magnetic field and the current
flow, the transport was found to be anisotropic, somewhat
reminiscent of the anisotropy observed at low temperature in high
mobility GaAs-based 2DEG \cite{Lilly99,Pan99}. Thus the N=1 LL
physics of our CdTe 2DEG looks globally similar to the one
observed in the well-known GaAs-based 2DEG, leaving open a
possible observation of the $\nu$=5/2 FQH state provided further
significant improvement are made in terms of sample quality.
Interestingly, the $\nu$=5/2 FQH state in our experimental
conditions (B=7.6 T, $g^{*}=-1.6$) would be associated with an
unprecedented Zeeman energy of $\sim 8$K. This could shed further
light on the issue of the spin polarization in this state which is
still a crucial point to validate its description using the Moore
Read wave function \cite{MooreRead} exhibiting exotic non-abelian
statistics.

In conclusion, we have shown that the 2DEG in a CdTe quantum well
can have a high quality, leading to the observation of pronounced
FQH states in the upper spin branch of the N=0 LL, as well as
emergent FQH minima in the N=1 LL. The physics of these FQH state
is strongly influenced by the intrinsic Zeeman energy, resulting
in the complete spin polarization of the FQH ground state, in
agreement with a CF approach for FQH effect. The high quality of
the 2D electron gas in CdTe offers a promising single valley
``model system'' to study delicate many-body effects in the
presence of a relatively high Zeeman energy. \acknowledgements{The
authors acknowledge the financial support from
EC-EuroMagNetII-228043, EC-ITEM MTKD-CT-2005-029671,
CNRS-PICS-4340, MNiSW- N20205432/1198 and
ERDF-POIG.01.01.02-00-008/08 grants.}

\end{document}